\documentclass[aps,prd,twocolumn,letterpaper,amsmath,showpacs,showkeys]{revtex4}

\newcommand{\CP}{\textit{CP}}
\newcommand{\half}{\frac{1}{2}}
\newcommand{\Vfac}{V\sub{fac}}
\newcommand{\sub}[1]{_{\mbox{\scriptsize#1}}}
\newcommand{\Inv}[1]{\frac{1}{#1}}
\newcommand{\hc}{^\dagger}
\newcommand{\Mg}{{\cal M}}
\newcommand{\diag}{\mathop{\textrm{diag}}\nolimits}
\newcommand{\MeV}{\,\textrm{MeV}}
\newcommand{\GeV}{\,\textrm{GeV}}
\newcommand{\degree}{^\circ}
\renewcommand{\Im}{\mathop{\textrm{Im}}\nolimits}
\renewcommand{\arraystretch}{1.3}

\begin{document}

\title{The Fritzsch Ansatz Revisited}
\author{W. K. Sze} 
\affiliation{Department of Physics, National Taiwan Normal University,
  Taipei, Taiwan 116} 
\date{\today}                                                

\begin{abstract}
\noindent
A modified Fritzsch ansatz for the quark mass matrices is proposed
to account for the hierarchical structure of the CKM~matrix.
To allow for the observed \CP~asymmetry, restrictions have to be imposed
on the relative phase degree of freedom among the weak eigenstates of the
quark fields, as for example by certain additional symmetries.
The ansatz can be accommodated in extensions to the Standard Model,
such as the multiple Higgs doublets models.
\end{abstract} 


\maketitle                                                   

\section{Introduction}

\noindent
It is an empirical fact that the left-handed quark mixing matrix
(the CKM matrix)~$V$\cite{CKM} differs from the unit matrix only slightly.
This is manifest in Wolfenstein's parameterization of~$V$:\cite{wolfenstein}
\begin{equation}
\label{ckm-matrix:wolf}
  V = \begin{pmatrix}
    1-\half\lambda^2 & \lambda               & A\lambda^3(\rho - i\eta) \\
    -\lambda            & 1 - \half\lambda^2 & A\lambda^2 \\
    A\lambda^3(1-\rho-i\eta) & -A\lambda^2      & 1
  \end{pmatrix},
\end{equation}
which is valid to leading orders of the small parameter $\lambda=0.220$.
Since it is known that the parameters $A$, $\rho$, and~$\eta$ are all of
order one, successive off-diagonal elements of~$V$ are of ascending orders
of~$\lambda$.
In this parameterization,
the \CP-violating phases are assigned to the elements $V_{ub}$ and~$V_{td}$,
with their phases being denoted as $-\gamma$ and~$-\beta$, respectively,
where $\beta$ and~$\gamma$ are angles appearing in the $d$-$b$
unitarity triangle.

One may speculate that this hierarchical structure of~$V$ has a deeper
meaning connected with the hierarchy of the quark masses, in which case
one should be able to put~$V$ in a form which reflects this connection.
An early attempt to establish such a relation was the ansatz proposed
by Fritzsch,\cite{fritzsch}
in which it is assumed that only the $b$ and~$t$ quarks have
non-zero bare masses, the effect of which are cascaded through
inter-generation couplings to give masses to the lighter quarks.
In the context of the Standard Model~(SM),
various phases in the mass matrices can be absorbed into the quark fields.
Hence one has to impose additional constraints among the quark phases,
in order that the ansatz be consistent with the observed \CP~asymmetry.
Even then, the numerical relations among the quark mass ratios and
mixing angles as predicted by the ansatz enjoy only a partial success.

In this paper, we will show that with certain modification the difficulty
of the ansatz can be circumvented,
while maintaining its prediction of a hierarchical CKM matrix.
To illustrate our point, let us put~$V$ in a factorized form,
expressing it as a product of successive rotations about the principal axes.
We notice that~$\Vfac$ as given below is of a form similar to
Eq.~(\ref{ckm-matrix:wolf}):
\begin{widetext}
\begin{align}
  \Vfac & =
    \begin{pmatrix}
      1-\half|k_u|^2\lambda^2 & k_u\lambda & 0 \\
      -k_u^*\lambda & 1-\half|k_u|^2\lambda^2 & 0 \\
      0 & 0 & 1
    \end{pmatrix}
    \begin{pmatrix}
      1 & 0 & 0 \\
      0 & 1 & A\lambda^2 \\
      0 & -A\lambda^2 & 1
    \end{pmatrix}
    \begin{pmatrix}
      1-\half|k_d|^2\lambda^2 & k_d\lambda & 0 \\
      -k_d^*\lambda & 1-\half|k_d|^2\lambda^2 & 0 \\
      0 & 0 & 1
    \end{pmatrix} \nonumber \\
  \label{ckm-matrix:fac}
    & \simeq \begin{pmatrix}
      1-\half|k|^2\lambda^2-i\delta   & k\lambda        & k_u A\lambda^3 \\
      -k^*\lambda       & 1 - \half|k|^2\lambda^2+i\delta & A\lambda^2 \\
      k_d^*A\lambda^3 & -A\lambda^2      & 1
    \end{pmatrix},
\end{align}
\end{widetext}
In the first and third factors of~$\Vfac$, phase factors are absorbed
in the complex parameters $k_u$ and~$k_d$, respectively.
On the second line we have dropped terms of orders higher than~$\lambda^2$
in the upper-left $2\times2$ block of~$\Vfac$,
consistent with what is done in Wolfenstein's
parameterization~(\ref{ckm-matrix:wolf}).
Furthermore we have denoted $k\equiv k_u+k_d$ and
$\delta\equiv-\half i(k_dk_u^*-k_d^*k_u)\lambda^2=\Im(k_dk_u^*)\,\lambda^2$.
Since $k_u$ and~$k_d$ are of order one (see below),
the contributions of~$\mp i\delta$ to the magnitudes~$|V_{ud}|$ and~$|V_{cs}|$
respectively cannot exceed~$O(\lambda^4)$.
If we ignore small phases of $V_{ud}$ and~$V_{cs}$,
as was done in Eq.~(\ref{ckm-matrix:wolf}),
then these $i\delta$ terms can be dropped as well.

If we choose $k=k_u+k_d$ to be unity,
which amounts to identifying the variables $\lambda$ and~$A$
in Eq.~(\ref{ckm-matrix:wolf}) and~(\ref{ckm-matrix:fac}),
then comparing the two equations we get $V=\Vfac$ with:
\begin{equation}
\label{ku-kd}
\begin{split}
k_u &= |k_u|\,e^{-i\gamma} = \rho-i\eta, \\
k_d &= |k_d|\,e^{i\beta} = 1-\rho+i\eta.
\end{split}
\end{equation}
If the factorized form~(\ref{ckm-matrix:fac}) of~$V$ is of any physical
significance,
it provides a hint to the quark mass matrices in the weak basis.
We will show that it is indeed consistent with a variant form
of the Fritzsch ansatz.

A brief review of the Fritzsch ansatz, as well as a recapitulation of its
short-comings, are given in the next section.
In Section~III we present a modified ansatz which is free of such
difficulties and would yield~$V$ as given by Eq.~(\ref{ckm-matrix:fac}).
In Section~IV the parameters in the quark mass matrix elements are estimated
from empirical data.
We give our conclusion in the final section.

\section{A Brief Review}

\noindent
In terms of the quark fields~$\Psi_U\equiv(u,c,t)^T$ and
$\Psi_D\equiv(d,s,b)^T$, the mass terms in the Lagrangian are:
\begin{equation}
\label{mass-lag}
{\cal L}_m = -(\bar\Psi_{U,L} M_U\Psi_{U,R} + \bar\Psi_{D,L} M_D\Psi_{D,R}) +
  \hbox{h.~c.},
\end{equation}
where as usual $\Psi_{I,L}$ and~$\Psi_{I,R}$ ($I=U,D$) denotes the left- and
right-handed fields $\half(1\mp\gamma_5)\Psi_I$, respectively.
If $\Psi_U$ and~$\Psi_D$ are expressed in the weak basis,
as is implicitly assumed above,
the mass matrices $M_U$ and~$M_D$ would in general not be diagonal.

The numerical coincidence $\sqrt{m_d/m_s} \simeq \lambda$ was noticed
in the early days.
As an endeavor to predict this relation,
Fritzsch proposed an ansatz for $M_U$ and~$M_D$.\cite{fritzsch}
Suppose these to be Hermitian matrices,
for two generations of quarks the ansatz reads:
\begin{equation}
\label{fritz-2}
M_U = \begin{pmatrix} 0 & A_U \\ A_U^* & B_U \end{pmatrix},\qquad
M_D = \begin{pmatrix} 0 & A_D \\ A_D^* & B_D \end{pmatrix},
\end{equation}
where $B_I$ ($I=U,D$) are real and positive.
If we assume that $|A_I|\ll B_I$, we would then obtain
\begin{equation}
\label{fritz-2:predict}
\lambda \simeq \sqrt{\frac{m_d}{m_s}} - \sqrt{\frac{m_u}{m_c}} \simeq 0.22,
\end{equation}
which fits well with the experimental value for~$\lambda$.

The interpretation for Eq.~(\ref{fritz-2}) is that only the heaviest quarks
have non-zero bare masses,
and there are couplings among quarks of neighboring generations.
The lighter quarks get their masses only by cascading effects through these
inter-generational couplings.
Accordingly the Fritzsch ansatz for three generations of quarks is:
\begin{equation}
\label{fritz-3}
  M_I = \begin{pmatrix}
    0 & A_I & 0 \\
    A_I^* & 0 & B_I \\
    0 & B_I^* & C_I
  \end{pmatrix}, \qquad \hbox{where~$I=U$ or~$D$.}
\end{equation}
Here we have $|A_I|\ll |B_I|\ll C_I$,
with the real parameters~$C_I$ assumed to be real and positive.

The ansatz~(\ref{fritz-3}) did give predictions in qualitative agreement
with empirical facts.
But as eventually more experimental data for the heavy quark masses
and mixing matrix elements such as $V_{ub}$ and~$V_{cb}$ are gathered,
it is realized that the success of the relation~(\ref{fritz-2:predict})
does not extend to the three-generation case.

Another problem is with the observed \CP-violation.
With so many vanishing elements in $M_U$ and~$M_D$,
one can normally utilize the phase degree of freedom of the quark fields
to put them in real forms,
and the resulting quark mixing matrix would have been real as well.
The observed \CP~asymmetry would then have to be accounted for by other
mechanism, which is hard to come by given the present experimental
data.\cite{cp-asym:expt}
Hence restrictions must be imposed on these phase ambiguities,
as was implicitly stated in Fritzsch original ansatz.\cite{fritzsch}

\section{A Modified Fritzsch Ansatz}

\noindent
Failure of the Fritzsch ansatz~(\ref{fritz-3}) can be traced to the
Hermitian nature of the mass matrices,
which is a very stringent condition on the magnitudes of the matrix elements.
A natural extension would then be relaxing the Hermiticity assumption for
$M_U$ and~$M_D$,
while retaining their general texture as given by Eq.~(\ref{fritz-3}).

As mentioned in the previous section,
in order to get a complex CKM~matrix,
we should require that $M_U$ and~$M_D$ be essentially complex,
i.~e., they cannot both be rendered real by a redefinition
of the phases of the quark weak eigenstates.
We assume that the only global phase symmetry among the quark weak
eigenstates is:
\begin{subequations}
\label{phase-sym}
\begin{align}
&\Psi_L \equiv \begin{pmatrix} \Psi_{U,L} \\ \Psi_{D,L} \end{pmatrix}
  \to \Psi_L' = \begin{pmatrix} P_U\Psi_{U,L} \\ P_D\Psi_{D,L} \end{pmatrix},
  \nonumber \\
&\Psi_{U,R} \to P_U\Psi_{U,R}, \qquad
\Psi_{D,R} \to P_D\Psi_{D,R},
\end{align}
where the diagonal phase matrices $P_U$ and~$P_D$ assume the following
restrictive form:
\begin{equation}
\label{phase-matrix}
\begin{split}
P_U &= \diag(e^{i\alpha}, e^{i\alpha'}, e^{i\alpha}), \\
P_D &= \diag(e^{i\chi}, e^{i\chi'}, e^{i\chi}).
\end{split}
\end{equation}
\end{subequations}
Briefly put, the transformation~(\ref{phase-sym}) is:
$\Psi_U\to P_U\Psi_U$, $\Psi_D\to P_D\Psi_D$.

In the~SM, there is a much richer phase degree of freedom for the quark fields
than that given by Eq.~(\ref{phase-sym}).
Its restriction to Eq.~(\ref{phase-sym}) therefore requires extensions to
the~SM.
One example would be the Two Higgs Doublets Model, in which
the up-type quarks get their masses from a scalar doublet~$\phi_1$,
while the down-type quarks get theirs from another scalar doublet~$\phi_2$,
and the two different sets of (quark and scalar) fields respect two distinct
$U(1)$ or higher symmetries,
with each field carrying non-trivial quantum number.
Each of the scalar doublets $\phi_1$ and~$\phi_2$ may conceivably be replaced
by a set of doublets, if the stringent constraints on FCNC can be met.
We will not elaborate upon these possibilities beyond the
requirement~(\ref{phase-sym}) in the following discussions.

\subsection{The Two-generation Case}

\noindent
To begin with,
we will illustrate our argument for the case of two generations of quarks.
Instead of Eq.~(\ref{fritz-2}), we try the following modified ansatz:
\begin{equation}
\label{ansatz-2}
M_U = \begin{pmatrix} 0 & A_U \\ A_U' & B_U \end{pmatrix},\qquad
M_D = \begin{pmatrix} 0 & A_D \\ A_D' & B_D \end{pmatrix}.
\end{equation}
We will assume that only $B_I$ ($I=U,D$) are real and positive,
while $A_I$ and~$A_I'$ are distinct complex numbers.
The Fritzsch-like condition $|A_I|, |A_I'| \ll B_I$ is still supposed to hold.
The Eq.~(\ref{phase-sym}) can be recast as the following transformation
on these mass matrix elements:
\begin{equation}
\label{phase-sym:2g}
\begin{split}
&A_U \to e^{i\Delta\alpha}A_U, \qquad
A_U' \to e^{-i\Delta\alpha}A_U', \\
&A_D \to e^{i\Delta\chi}A_D, \qquad
A_D' \to e^{-i\Delta\chi}A_D', \\
&\mbox{$B_U$ and~$B_D$ unchanged.}
\end{split}
\end{equation}
Here we have denoted $\Delta\alpha=\alpha'-\alpha$ and $\Delta\chi=\chi'-\chi$.
In general we need to carry out bi-unitary transformations on $M_U$ and~$M_D$
to put them in diagonal form.

Such diagonalization procedures are well known.
We will nonetheless briefly re-state the steps to fix our phase convention.
We know that, any $2\times2$ Hermitian matrix
\begin{displaymath}
  \Mg = \begin{pmatrix} \Mg_{11} & \Mg_{12} \\
    \Mg_{12}^* & \Mg_{22} \end{pmatrix}
\end{displaymath}
can be diagonalized $\Mg\to\Mg_{\rm diag}\equiv O\hc\Mg O$ by a uni-modular
unitary matrix:
\begin{equation}
\label{Omatrix}
  O = \begin{pmatrix}\cos\theta & e^{i\varphi}\sin\theta \\
    -e^{-i\varphi}\sin\theta & \cos\theta \end{pmatrix}.
\end{equation}
The real parameters $\theta$ and~$\varphi$ can be expressed
in terms of elements of~$\Mg$:
\begin{equation}
  \tan2\theta = \frac{2|\Mg_{12}|}{\Mg_{22}-\Mg_{11}}, \qquad
  \varphi = \arg\Mg_{12},
\end{equation}
from the assumption that $O\hc\Mg O$ is diagonal.

For~$M=M_U$ or~$M_D$ as given by Eq.~(\ref{ansatz-2}),
the transformation matrices which bi-diagonalize~$M$ are those
which diagonalize the Hermitian matrices $MM\hc$ and $M\hc M$.
Take~$M_D$ for example.  We have
\begin{align*}
M_DM_D\hc &= \begin{pmatrix}
  |A_D|^2 & A_DB_D \\ A_D^*B_D & |A_D'|^2+B_D^2 \end{pmatrix}, \\
M_D\hc M_D &= \begin{pmatrix}
  |A_D'|^2 & A_D^{\prime*}B_D \\ A_D'B_D & |A_D|^2+B_D^2 \end{pmatrix}.
\end{align*}
We assume that they are diagonalized by the unitary matrices $L_D$ and~$R_D$,
respectively,
where $L_D$ and~$R_D$ are both of forms similar to Eq.~(\ref{Omatrix}):
\begin{equation}
\label{LR-matrix:0}
\begin{split}
  L_D &= \begin{pmatrix} \cos\theta_d & e^{i\varphi_d}\sin\theta_d \\
    -e^{-i\varphi_d}\sin\theta_d & \cos\theta_d \end{pmatrix}, \\
  R_D &= \begin{pmatrix} \cos\theta_d' & e^{i\varphi_d'}\sin\theta_d' \\
    -e^{-i\varphi_d'}\sin\theta_d' & \cos\theta_d' \end{pmatrix}.
\end{split}
\end{equation}
Bearing in mind the Fritzsch condition $|A_D|, |A_D'| \ll B_D$,
we obtain the expressions for the parameters in $L_D$ and~$R_D$:
\begin{equation}
\label{mixing-parameters:left-right}
\begin{split}
\theta_d &\simeq \frac{|A_D|}{B_D}, \qquad \varphi_d = +\arg A_D; \\
\theta_d' &\simeq \frac{|A_D'|}{B_D}, \qquad \varphi_d' = -\arg A_D'.
\end{split}
\end{equation}
The matrix $L_D\hc M_DR_D$ will be diagonal as desired,
but the lighter mass eigenvalue $-A_DA_D'/B_D$ will in general be complex.
It is desirable to absorb the extra phases in the transformation matrices
$L_D$ and~$R_D$ so that all the mass eigenvalues are real and positive.
Thus we take, instead of $L_D$ and~$R_D$,
the following transformation matrices:
\begin{equation}
\label{LR-matrices:1a}
\begin{split}
\tilde L_D & \equiv
L_D \begin{pmatrix} e^{i\varphi_d} & 0 \\ 0 & 1 \end{pmatrix}
  = \begin{pmatrix} e^{i\varphi_d}\cos\theta_d & e^{i\varphi_d}\sin\theta_d \\
    -\sin\theta_d & \cos\theta_d \end{pmatrix}, \\
\tilde R_D & \equiv
R_D \begin{pmatrix} -e^{i\varphi_d'} & 0 \\ 0 & 1 \end{pmatrix}
  = \begin{pmatrix} -e^{i\varphi_d'}\cos\theta_d' &
      e^{i\varphi_d'}\sin\theta_d' \\
    \sin\theta_d' & \cos\theta_d' \end{pmatrix}.
\end{split}
\end{equation}
With $\tilde L_D$ and~$\tilde R_D$ as given above,
$M_D$ is transformed to its diagonal form~$\hat M_D$:
\begin{displaymath}
\hat M_D = \tilde L_D\hc M_D\tilde R_D.
\end{displaymath}
Bearing in mind that from Eq.~(\ref{mixing-parameters:left-right})
we have relations such as $e^{-i\varphi_d}A_D=|A_D|$, etc.,
and keeping only leading terms of the small parameters $\theta_d$
and~$\theta_d'$, one verifies that $\hat M_D$ is diagonal:
\begin{equation}
\label{M-diag}
\hat M_D \simeq \begin{pmatrix} \displaystyle\frac{|A_DA_D'|}{B_D} & 0 \\
    0 & B_D \end{pmatrix}
  = \begin{pmatrix} m_d & 0 \\ 0 & m_s \end{pmatrix}.
\end{equation}
The corresponding down-quark mass eigenstates are then given by:
\begin{equation}
\label{dquark-mass-eigenstates}
\hat\Psi_{D,L} = \tilde L_D\hc\Psi_{D,L}, \qquad
\hat\Psi_{D,R} = \tilde R_D\hc\Psi_{D,R},
\end{equation}
where $\Psi_{D,L}$ and $\Psi_{D,R}$ are the weak eigenstates appearing in
Eq.~(\ref{mass-lag}).

The Cabibbo matrix~$V_2$, which measures the mis-alignment between the
up- and down-type left-handed quark eigenstates,
would then be given as:
\begin{align}
\label{Cabibbo-matrix}
V_2 &= \tilde L_U\hc\tilde L_D \nonumber \\
  &= \begin{pmatrix} c_u e^{-i\varphi_u} & -s_u \\
      s_u e^{-i\varphi_u} & c_u \end{pmatrix}
    \begin{pmatrix} c_d e^{i\varphi_d} & s_d e^{i\varphi_d} \\
      -s_d & c_d \end{pmatrix},
\end{align}
where as usual the abbreviations $c_u\equiv \cos\theta_u$,
$c_d\equiv \cos\theta_d$, etc., have been used.
We will make use of the phase degree of freedom as given by
Eq.~(\ref{phase-sym}) or~(\ref{phase-sym:2g}) to adjust the phase of $A_U$
or~$A_D$ to get $\varphi_d=\varphi_u$.
Then we have
\begin{equation}
\label{Cabibbo-matrix:2}
V_2 = \begin{pmatrix}
  \cos\theta & \sin\theta \\ -\sin\theta & \cos\theta \end{pmatrix},
\end{equation}
where $\theta\equiv \theta_d-\theta_u$.
Hence $V_2$ can be put in a real form,
as it should with two generations of quarks.

Relations between the mixing angle~$\theta$ and the quark mass ratios
can be deduced from the preceding equations.
For illustration we set $m_u\simeq0$, in which case we have $\theta=\theta_d$.
From Eq.~(\ref{mixing-parameters:left-right}) and~(\ref{M-diag}) we get
\begin{displaymath}
\biggl(\frac{m_d}{m_s}\biggr)^{\!\half} = \frac{\sqrt{|A_DA_D'|}}{B_D},
  \qquad \hbox{but}\quad \theta = \frac{|A_D|}{B_D}.
\end{displaymath}
In the traditional Fritzsch scheme, we have $A_D=A_D'$,
and the equality~$\theta=\sqrt{m_d/m_s}$ results.
Although for the $d$-$s$ pair this is a success,
such success does not survive in the three-generation case,
and the Fritzsch ansatz fails consequently.

On the other hand, we see that in this modified ansatz, the mixing angle
and the square-root of the mass ratio need not be equal anymore,
and the data can again be fitted.
The price to pay is that now the model loses predictive power.

\subsection{The Three-generation Case}

\noindent
We next proceed to the three generation case.
Our modified ansatz for the mass matrices reads:
\begin{equation}
\label{mass-matrices}
M_U = \begin{pmatrix} 0 & A_U & 0 \\
  A_U' & 0 & B_U \\
  0 & B_U' & C_U \end{pmatrix},\quad
M_D = \begin{pmatrix} 0 & A_D & 0 \\
  A_D' & 0 & B_D \\
  0 & B_D' & C_D \end{pmatrix}.
\end{equation}
Again we will assume that only $C_I$~($I=U,D$) are real and positive;
furthermore we have $|A_I|,|A_I'|\ll |B_I|,|B_I'|\ll C_I$.
The phase transformation~(\ref{phase-sym}) can be re-stated as:
\begin{equation}
\label{phase-sym:3g}
\begin{split}
(A_U, B_U') &\to e^{i\Delta\alpha}(A_U, B_U'), \\
(A_U', B_U) &\to e^{-i\Delta\alpha}(A_U', B_U), \\
(A_D, B_D') &\to e^{i\Delta\chi}(A_D, B_D'), \\
(A_D', B_D) &\to e^{-i\Delta\chi}(A_D', B_D), \\
\mbox{$C_U$ and}~&\mbox{$C_D$ unchanged.}
\end{split}
\end{equation}
Here $\Delta\alpha=\alpha'-\alpha$, etc.,
as in the previous subsection.

These mass matrices are diagonalized in two stages.
Again take~$M_D$ as example.  Let:
\begin{equation}
\label{L2D-matrix}
  \tilde L_{2D} = \begin{pmatrix} 1 & 0 & 0 \\
    0 & e^{i\varphi_{2d}}\cos\theta_{2d} & e^{i\varphi_{2d}}\sin\theta_{2d} \\
    0 & -\sin\theta_{2d} & \cos\theta_{2d} \end{pmatrix},
\end{equation}
wherein the parameters are
\begin{equation}
\label{mixing-parameters:2d}
\theta_{2d} = \frac{|B_D|}{C_D}, \qquad \varphi_{2d} = +\arg B_D.
\end{equation}
The right-handed quark transformation matrix~$\tilde R_{2D}$ is similarly
defined,
its parameters being $\theta_{2d}'=|B_D'|/C_D$ and~$\varphi_{2d}'=-\arg B_D'$.
With $\tilde L_{2D}$ and~$\tilde R_{2D}$, the matrix~$M_D$ is first transformed
to~$M_D'$, in which the lower-right $2\times2$ block is diagonal:
\begin{align}
M_D' &\equiv \tilde L_{2D}\hc M_D\tilde R_{2D} \nonumber \\
\label{intermediate-diag}
  &\simeq \begin{pmatrix} 0 & -e^{i\varphi_{2d}'}A_D & 0 \\
    e^{-i\varphi_{2d}}A_D' & \displaystyle\frac{|B_DB_D'|}{C_D} & 0 \\
    0 & 0 & C_D \end{pmatrix},
\end{align}
where terms in higher orders of the small parameters $\theta_{2d}$
and~$\theta_{2d}'$ have been dropped.

In the subsequent diagonalization of $M_D'$,
we could have used the matrices $\tilde L_{1D}$ and~$\tilde R_{1D}$,
which effect a rotation between the first and second generations,
but are otherwise defined similarly as $\tilde L_{2D}$ and~$\tilde R_{2D}$:
\begin{displaymath}
\hat M_D \equiv \tilde L_{1D}\hc M_D'\tilde R_{1D}.
\end{displaymath}
But since we would like to have the phase assignment of the final CKM matrix
agreeing with Wolfenstein's,
as given by Eq.~(\ref{ckm-matrix:wolf}) or~(\ref{ckm-matrix:fac}),
we will redefine the transformation matrices as follows:
\begin{displaymath}
\tilde L_{1D} \to L_{1D}\equiv \tilde L_{1D}P_{1D}, \qquad
\tilde R_{1D} \to \bar R_{1D}\equiv \tilde R_{1D}P_{1D},
\end{displaymath}
where the phase matrix~$P_{1D}\equiv \diag(e^{-i\varphi_{1d}}, 1, 1)$,
where $\varphi_{1d}$ is defined in Eq.~(\ref{mixing-parameters:1d}) below.
Hence $L_{1D}$ (but not $\bar R_{1D}$) is uni-modular.
The final diagonalized~$\hat M_D$ is not altered if we use
$L_{1D}$ and~$\bar R_{1D}$ instead of $\tilde L_{1D}$ and~$\tilde R_{1D}$.
For the pair $L_{1D}$ and~$\tilde L_{1D}$, the above procedure is similar to
the reverse of the redefinition from Eq.~(\ref{LR-matrix:0})
to~(\ref{LR-matrices:1a}) in the previous subsection.
The explicit form for~$L_{1D}$ is:
\begin{equation}
\label{L1D-matrix}
L_{1D} = 
  \begin{pmatrix} \cos\theta_{1d} & e^{i\varphi_{1d}}\sin\theta_{1d} & 0 \\
    -e^{-i\varphi_{1d}}\sin\theta_{1d} & \cos\theta_{1d} & 0 \\
    0 & 0 & 1 \end{pmatrix},
\end{equation}
From Eq.~(\ref{intermediate-diag}),
and recalling that $\varphi_{2d}'=-\arg B_D'$,
we see that the parameters in~$L_{1D}$ are given by
\begin{equation}
\label{mixing-parameters:1d}
\theta_{1d} = \frac{|A_DC_D|}{|B_DB_D'|}, \qquad
\varphi_{1d} = \arg\frac{A_D}{B_D'}-\pi.
\end{equation}
The diagonal mass matrix~$\hat M_D$ would then be
\begin{subequations}
\label{dquark-masses}
\begin{align}
\hat M_D &\equiv L_{1D}\hc M_D'\bar R_{1D} \nonumber \\
  &\simeq \begin{pmatrix}
    \displaystyle\frac{|A_DA_D'|\,C_D}{|B_DB_D'|} & 0 & 0 \\
    0 & \displaystyle\frac{|B_DB_D'|}{C_D} & 0 \\
    0 & 0 & C_D \end{pmatrix}.
\end{align}
Again terms of higher order in the small parameters $\theta_{1d}$
and~$\theta_{1d}'$ have been dropped.

The two-step process described above amounts to the following transformation
of $M_D$ to~$\hat M_D$:
\begin{align}
\hat M_D &= L_{1D}\hc\tilde L_{2D}\hc M_D\tilde R_{2D}\bar R_{1D}
    \nonumber \\
  &= \diag(m_d, m_s, m_b).
\end{align}
\end{subequations}
Correspondingly the down-type quark mass eigenstates are
$\hat\Psi_{D,L}=L_{1D}\hc\tilde L_{2D}\hc\Psi_{D,L}$ and
$\hat\Psi_{D,R}=\bar R_{1D}\hc\tilde R_{2D}\hc\Psi_{D,R}$.

Exactly parallel procedures apply in the diagonalization of~$M_U$.
With similar notations,
the transformation of~$M_U$ to its diagonal form~$\hat M_U$ reads:
\begin{align}
\hat M_U &\equiv L_{1U}\hc\tilde L_{2U}\hc M_U\tilde R_{2U}\bar R_{1U}
    \nonumber \\
\label{uquark-masses}
  &= \begin{pmatrix} \displaystyle\frac{|A_UA_U'|\,C_U}{|B_UB_U'|} & 0 & 0 \\
    0 & \displaystyle\frac{1}{C_U}|B_UB_U'| & 0 \\
    0 & 0 & C_U \end{pmatrix} \nonumber \\
  &= \diag(m_u, m_c, m_t).
\end{align}
The matrices~$L_{1U}$, etc., are defined similar to
their counterparts for the down-type quarks.
For example, the explicit form of~$L_{1U}$ is:
\begin{equation}
\label{L1U-matrix}
L_{1U} =
  \begin{pmatrix} \cos\theta_{1u} & e^{i\varphi_{1u}}\sin\theta_{1u} & 0 \\
    -e^{-i\varphi_{1u}}\sin\theta_{1u} & \cos\theta_{1u} & 0 \\
    0 & 0 & 1 \end{pmatrix}.
\end{equation}
where the parameters $\theta_{1u}$ and~$\varphi_{1u}$ are defined similarly
as in Eq.~(\ref{mixing-parameters:1d}), with appropriate change of suffices.
Likewise~$L_{2U}$ is given by:
\begin{equation}
\label{L2U-matrix}
\tilde L_{2U} = \begin{pmatrix} 1 & 0 & 0 \\
    0 & e^{i\varphi_{2u}}\cos\theta_{2u} & e^{i\varphi_{2u}}\sin\theta_{2u} \\
    0 & -\sin\theta_{2u} & \cos\theta_{2u} \end{pmatrix},
\end{equation}
with $\theta_{2u}$ and~$\varphi_{2u}$ defined similarly
as in Eq.~(\ref{mixing-parameters:2d}).
The up-type quark mass eigenstates would then be given by
$\hat\Psi_{U,L}=L_{1U}\hc\tilde L_{2U}\hc\Psi_{U,L}$ and
$\hat\Psi_{U,R}=\bar R_{1U}\hc\tilde R_{2U}\hc\Psi_{U,R}$.

For evaluation of the CKM matrix,
only the left-handed quark fields need be considered.
For these fields, the expression of the weak eigenstates in terms of the
mass eigenstates are:
\begin{align*}
\Psi_{U,L} &= \tilde L_{2U}L_{1U}\hat\Psi_{U,L}, \\
\Psi_{D,L} &= \tilde L_{2D}L_{1D}\hat\Psi_{D,L}.
\end{align*}
The CKM matrix is thus given by
\begin{equation}
\label{ckm-fritz:3}
V = L_{1U}\hc\tilde L_{2U}\hc\tilde L_{2D}L_{1D}.
\end{equation}
From the expression for $\tilde L_{2D}$ and~$\tilde L_{2U}$ as shown in
Eq.~(\ref{L2D-matrix}) and~(\ref{L2U-matrix}),
we get the middle factor on the right hand side of this equation:
\begin{multline}
\label{Cabibbo-matrix:3g}
\tilde L_{2U}\hc\tilde L_{2D} = \\
  \begin{pmatrix} 1 & 0 & 0 \\
    0 & c_{2u} e^{-i\varphi_{2u}} & -s_{2u} \\
    0 & s_{2u} e^{-i\varphi_{2u}} & c_{2u} \end{pmatrix}
  \begin{pmatrix} 1 & 0 & 0 \\
    0 & c_{2d} e^{i\varphi_{2d}} & s_{2d} e^{i\varphi_{2d}} \\
    0 & -s_{2d} & c_{2d} \end{pmatrix},
\end{multline}
where the abbreviations $c_{2u}\equiv \cos\theta_{2u}$,
$s_{2u}\equiv \sin\theta_{2u}$, etc., have been used.
We then make use of the phase ambiguity as given by
Eq.~(\ref{phase-sym}) or~(\ref{phase-sym:3g}) to adjust the phase of $B_U$
or~$B_D$, so that we have $\varphi_{2d}=\varphi_{2u}$.
One then verifies that Eq.~(\ref{Cabibbo-matrix:3g}) reduce to the
following real form:
\begin{equation}
\label{V2-matrix}
\tilde L_{2U}\hc\tilde L_{2D} = \begin{pmatrix} 1 & 0 & 0 \\
  0 &  \cos\theta_2 & \sin\theta_2 \\
  0 & -\sin\theta_2 & \cos\theta_2 \end{pmatrix},
\end{equation}
where we have denoted $\theta_2\equiv \theta_{2d}-\theta_{2u}$.
Substituting Eq.~(\ref{L1D-matrix}), (\ref{L1U-matrix}), and~(\ref{V2-matrix})
into Eq.~(\ref{ckm-fritz:3}), we finally get:
\begin{widetext}
\begin{equation}
\label{ckm-matrix:model}
V = \begin{pmatrix}
  c_{1u} & -s_{1u} e^{i\varphi_{1u}} & 0 \\
  s_{1u} e^{-i\varphi_{1u}} & c_{1u} & 0 \\
  0 & 0 & 1
\end{pmatrix}
\begin{pmatrix}
  1 & 0 & 0 \\
  0 &  c_2 & s_2 \\
  0 & -s_2 & c_2
\end{pmatrix}
\begin{pmatrix}
   c_{1d} & s_{1d} e^{i\varphi_{1d}} & 0 \\
  -s_{1d} e^{-i\varphi_{1d}} & c_{1d} & 0 \\
  0 & 0 & 1
\end{pmatrix},
\end{equation}
\end{widetext}
where again we have denoted $c_{1u}\equiv \cos\theta_{1u}$,
$s_{1u}\equiv \sin\theta_{1u}$, etc., for brevity.

We see that $V$ as given above would be equal to~$\Vfac$
in Eq.~(\ref{ckm-matrix:fac}), if we make the following identification:
\begin{align}
k_d\lambda &= e^{i\varphi_{1d}}\sin\theta_{1d}, \qquad
k_u\lambda = -e^{i\varphi_{1u}}\sin\theta_{1u}, \nonumber \\
\label{kvars}
A\lambda^2 &= \sin\theta_2 \equiv \sin(\theta_{2d}-\theta_{2u}).
\end{align}
The relations listed above would be `natural',
yielding values of $|k_u|$, $|k_d|$, and~$A$ of order unity,
if the angles $\theta_{1u}$, $\theta_{1d}$, $\theta_{2u}$, and $\theta_{2d}$
are all analytic functions of~$\lambda$ with leading~$O(\lambda)$ term, so that
$\sin(\theta_{2d}-\theta_{2u})\simeq \theta_{2d}-\theta_{2u}=O(\lambda^2)$.

If we compare $k_u$ and~$k_d$ as given by Eq.~(\ref{ku-kd}) and~(\ref{kvars}),
we get $\beta=\varphi_{1d}$ and $\gamma=-\varphi_{1u}-\pi$, modulo~$2\pi$.
The phase~$\varphi_{1d}$ and~$\varphi_{1u}$ are given by
Eq.~(\ref{mixing-parameters:1d}) and its counterpart for the up-type quarks.
Hence we get, modulo~$2\pi$:
\begin{equation}
\label{V:phases}
\beta = \arg\frac{A_D}{B_D'} - \pi, \qquad
\gamma = -\arg\frac{A_U}{B_U'}.
\end{equation}
We see that, in this ansatz, the \CP~violating phases $\beta$ and~$\gamma$
are determined by the phase mismatches between $A_D$ and~$B_D'$,
and between $B_U'$ and~$A_U$, respectively.
Note that the ratios $A_U/B_U'$ and $A_D/B_D'$ are invariant under the
transformation~(\ref{phase-sym:3g}).

\section{The Quark Mass Matrices}

\noindent
As mentioned in the previous section, empirical data indicate that
the mixing angles~$\theta_{1d}$, $\theta_{2d}$, etc., are of order~$\lambda$.
From Eq.~(\ref{mixing-parameters:2d}), (\ref{mixing-parameters:1d}), etc.,
we would then deduce that both $|B_D|/C_D$ and $|A_DC_D|/|B_DB_D'|$ are
of order~$\lambda$.
Similar conclusions hold for elements of~$M_U$.
Hence we rewrite the Fritzsch ansatz~(\ref{mass-matrices}) in the following
alternative form,
where appropriate powers of~$\lambda$ are extracted from the matrix elements:
\begin{equation}
\label{mass-matrices:red}
\begin{split}
M_U &= m_t \begin{pmatrix} 0 & a_u\lambda^3 & 0 \\
  a_u'\lambda^3 & 0 & b_u\lambda \\
  0 & b_u'\lambda & 1 \end{pmatrix}, \\
M_D &= m_b \begin{pmatrix} 0 & a_d\lambda^3 & 0 \\
  a_d'\lambda^3 & 0 & b_d\lambda \\
  0 & b_d'\lambda & 1 \end{pmatrix}.
\end{split}
\end{equation}
Presumably the complex parameters~$a_d$, etc., should be of order unity.
But as we will see below, most parameters in~$M_U$ are significantly smaller.
Hence Eq.~(\ref{mass-matrices:red}) should only be taken as a formal expansion.

Substituting Eq.~(\ref{mass-matrices:red}) into relations such as
Eqs.~(\ref{mixing-parameters:2d}) and~(\ref{mixing-parameters:1d}), we have
$\theta_{2d}=|b_d|\,\lambda$ and $\theta_{1d}=|a_d/b_db_d'|\,\lambda$,
and likewise for $\theta_{2u}$ and~$\theta_{1u}$.
Hence we get, from Eq. (\ref{kvars}) and~(\ref{V:phases}):
\begin{equation}
\label{kvars-red}
\begin{split}
k_u &= |k_u|\,e^{-i\gamma} = \Inv{|b_u|}\frac{a_u}{b_u'}, \\
k_d &= |k_d|\,e^{i\beta} = -\Inv{|b_d|}\frac{a_d}{b_d'}, \\
A\lambda &= |b_d| - |b_u|.
\end{split}
\end{equation}
The parameters $a_u'$ and~$a_d'$ are not constrained by the experimental data
on elements of~$V$.

The quark masses provide another set of data from which the parameters
can be fitted.
Substituting Eq.~(\ref{mass-matrices:red}) into Eq.~(\ref{dquark-masses})
and~(\ref{uquark-masses}),
we get the quark mass ratios in terms of the parameters~$a_u$, etc.:
\begin{equation}
\label{quark-mass-ratios}
\begin{split}
\frac{m_u}{m_c} &= \frac{|a_ua_u'|}{|b_ub_u'|^2}\,\lambda^2,\qquad
\frac{m_c}{m_t} = |b_ub_u'|\,\lambda^2, \\
\frac{m_d}{m_s} &= \frac{|a_da_d'|}{|b_db_d'|^2}\,\lambda^2,\qquad
\frac{m_s}{m_b} = |b_db_d'|\,\lambda^2.
\end{split}
\end{equation}

From Eq.~(\ref{kvars-red}) and~(\ref{quark-mass-ratios}),
values of $a_u$, $a_d$, etc., can be fitted from experimental data.
The parameter~$\lambda$ is known to a high precision:
\begin{equation}
\label{lambda}
\lambda = 0.220.
\end{equation}
The other parameters in~$V$ have also been determined,
albeit with less accuracies:\cite{PDG}
\begin{equation}
\label{mixing-parameters}
A = 0.86,\quad \rho = 0.20,\quad \eta = 0.34.
\end{equation}
Among these, the relative uncertainties of~$A$ is about~5\%,
while those of $\rho$ and~$\eta$ are significantly higher,
amounting to~30\% or more.

Among the input data used for fitting the values of $\rho$ and~$\eta$ above
is the value of $\sin2\beta\simeq0.74$ obtained in recent $B^0$-$\bar B^0$
experiments.\cite{cp-asym:expt}
As a result we can write $\beta\simeq23\degree$.
Our knowledge on the other angle~$\gamma$ is less precise,
the central value being $\gamma\sim60\degree$.\cite{PDG}

We will adopt the central values of the quark masses as given by the
Particle Data Group.
For the light quarks, the running mass values at the scale~$\mu_0\equiv 2\GeV$
are listed:\cite{PDG}
\begin{align}
m_u(\mu_0) &= 3.3\MeV, \qquad
m_d(\mu_0)  = 7.0\MeV, \nonumber \\
\label{light-quark-masses}
m_s(\mu_0) &= 120\MeV.
\end{align}
These are the current-algebra quark masses appearing in the mass Lagrangian.
They have typically rather high uncertainties,
ranging from 30\% to more than~50\%.

For the heavy quarks~$q=c,b,t$, on the other hand, the values quoted are the
physical masses~$\hat m_q$, which are defined by
$m_q(\mu=\hat m_q)=\hat m_q$:\cite{PDG}
\begin{align*}
\hat m_c &= 1.2\GeV, \qquad
\hat m_b  = 4.2\GeV, \\
\hat m_t &= 174\GeV.
\end{align*}
Since relations such as Eq.~(\ref{quark-mass-ratios}) are to be interpreted
at some fixed energy scale,
the heavy quark masses have to be run to the same scale~$\mu_0=2\GeV$
before they are substituted in Eq.~(\ref{quark-mass-ratios}).
We will take $\Lambda\sub{QCD}=220\MeV$,\cite{PDG}
and the heavy quark masses at~$\mu_0$ are:
\begin{align}
m_c(\mu_0) &= 1.06\GeV, \qquad
m_b(\mu_0)  = 4.83\GeV, \nonumber \\
\label{heavy-quark-masses}
m_t(\mu_0) &= 306\GeV.
\end{align}
The uncertainties of the physical heavy-quark masses range from a few percent
to around~15\%.
Nonetheless, taking into account that $\Lambda\sub{QCD}$
is only known to around~10\% precision, the uncertainties of the values
given by Eq.~(\ref{heavy-quark-masses}) can be comparable to those of the
light quark masses.

\begin{table}
\caption{\label{mass-parameters}Estimates of or constraints on parameters
appearing in $M_U$ and~$M_D$ as given by Eq.~(\ref{mass-matrices:red}).}
\begin{ruledtabular}
\renewcommand{\arraystretch}{1.6}
\begin{tabular}%
{r@{\extracolsep{0pt}}l@{\extracolsep{4em}}r@{\extracolsep{0pt}}l%
@{\extracolsep{2em}}l}
$|a_u|$& ${}\sim0.03$,& $|a_d|$& ${}\sim0.5$,& \\
$|a_u'|$& ${}\sim0.01$,& $|a_d'|$& ${}\sim0.8$,& \\
$|b_ub_u'|$& ${}\sim0.07$,& $|b_db_d'|$& ${}\sim0.6$,& \\
\noalign{\smallskip}
$\displaystyle\arg\frac{b_u'}{a_u}$& ${}\sim 1.0$,&
$\displaystyle\arg\frac{b_d'}{a_d}$& ${}\sim 2.7$,& (mod $2\pi$)\\
\noalign{\smallskip}
\multicolumn{4}{c}{$|b_d|-|b_u| \simeq 0.19$.}&
\end{tabular}
\end{ruledtabular}
\end{table}

From the above considerations, estimates of $|a_u|$, $|a_d|$, etc.,
as well as other constraints on these parameters can be deduced.
These are listed in Table~\ref{mass-parameters}.
From discussions on the uncertainties of the mixing matrix elements and
quark masses in Eq.\ (\ref{mixing-parameters})--(\ref{heavy-quark-masses}),
it can be inferred that typical values given in the table are also
subject to variation of a few tens of percent.

From the table we see that $|a_d|$ and~$|a_d'|$ are of order one, and it is
reasonable to assume that $|b_d|$ and~$|b_d'|$ are also of order unity.
Since $|b_u|$ are close in value to~$|b_d|$, it is also of this order.
On the other hand, the absolute values of the other parameters $|b_u'|$,
$|a_u|$, and~$|a_u'|$ in~$M_U$ are significantly smaller, reflecting the
larger discrepancy in mass values of the $u$, $c$, and~$t$ quarks.

\section{Conclusion}

\noindent
The observed CKM matrix~$V$ is close to the identity matrix, and has a neat
hierarchical form as manifested in Wolfenstein's parameterization.
One can accept the fact as it is, or try to explore for any possible
structures behind this pattern.

In this paper a modified Fritzsch ansatz for the quark mass matrices,
which is motivated by the factorized form of~$V$ as given by
Eq.~(\ref{ckm-matrix:fac}), is proposed.
The complex matrices $M_U$ and~$M_D$ have the same general texture
of those in the original Fritzsch ansatz,
but are neither symmetric nor Hermitian.
It was shown that, if $M_U$ and~$M_D$ have the hierarchical pattern as given
by~Eq.(\ref{mass-matrices:red}),
with values of parameters as listed on Table~\ref{mass-parameters},
the resulting mixing matrix~$V$ is in agreement with the observed one.

We see from Table~\ref{mass-parameters} that $|a_d|=O(1)$,
and it is also plausible that $|b_d|,|b_d'|=O(1)$.
From Eq.~(\ref{kvars-red}) it then follows naturally that $|k_d|$ is of
order unity.
The fact that $|A|=O(1)$ is also easily acceptable, from the same equation,
by noting that the two order-one parameters $b_d$ and~$b_u$ have a
difference in magnitudes of order~$\lambda$.
On the other hand, the fact that $|k_u|=|1-k_d|=\sqrt{\rho^2+\eta^2}=O(1)$
depends on the two small parameters $|a_u|,|b_u'|=O(10^{-2})$ having a ratio
close to one.
This may seem accidental,
but we have noted that the phase of $a_u/b_u'$ is invariant under
the transformation~(\ref{phase-sym}), and it is conceivable that there is
some related mechanism which ensures that the magnitude~$|a_u/b_u'|$
will not become too small, even when both $|a_u|$ and~$|b_u'|$ do.

Within the context of the~SM, one can utilize the phase degree of freedom
of the weak eigenstates of the quark fields to remove all phases in
$M_U$ and~$M_D$ as given by Eq.~(\ref{mass-matrices}) in the first place.
It follows then that, in a certain sense, the ansatz mandates new physics
beyond the~SM, in which the quark phase ambiguities is more restrictive
as in Eq.~(\ref{phase-sym}).
As mentioned in the beginning of Section~III, natural candidates which
come to one's mind are the multiple Higgs doublets models, although
other possibilities should not be ruled out.

\end{document}